# Comparison of the 3-fluid dynamic model with experimental data

**V.A. Kizka**

*Department of experimental nuclear physics,*
*V.N.Karazin Kharkiv National University, Kharkiv, 61022, Ukraine*
*E-mail: valeriy.kizka@karazin.ua*

**Abstract.** The article considers a way to compare large bulks of experimental data with theoretical calculations, in which the quality of theoretical models is clearly demonstrated graphically. The main idea of the method consists in grouping physical observables, represented by experiment and theoretical calculation, into samples, each of which characterizes a certain physical process. A further choice of a convenient criterion for comparing measurements and calculations, its calculation and averaging within each sample and then over all samples, makes it possible to choose the best theoretical model in the entire measurement area. Modern analysis of experimental data and their comparison with calculations in the search for exotic states of nuclear matter, where a huge amount of material has been accumulated over several decades, is still largely carried out by eye. Published theoretical data of the three-fluid dynamic model (3FD) applied to the experimental data from heavy-ion collisions at the energy range $\sqrt{s_{NN}}$ = 2.7 - 63 GeV are used as example of application of the developed methodology. When analyzing the results, the quantum nature of the fireball, created at heavy ion collisions, was taken into account. Thus, even at energy $\sqrt{s_{NN}}$ = 63 GeV of central collisions of heavy ions, there is a nonzero probability of fireball formation without ignition of the quark-gluon plasma (QGP). At the same time, QGP ignition at central collision energies above at least $\sqrt{s_{NN}}$ = 12 GeV occurs through two competing processes, through a first-order phase transition and through a smooth crossover. That is, in nature, these two possibilities are realized, which occur with approximately the same probabilities. Modern experiment and theory not only does not consider a fireball, born in collisions of relativistic heavy ions, as a quantum object that can have different quantum states with the same energy pumped into it, but the processing of experimental data itself does not provide for the imposition of any triggers to separate these states from each other. This work takes into account the quantum nature of the fireball and the need to analyze all the information accumulated over decades at once.

## Introduction

Articles devoted to the study of the formation of Quark-Gluon Plasma (QGP) in heavy ion collisions contain a huge amount of experimental and theoretical material [1–10]. If some criterion is used to assess the quality of the description of experimental data by some theoretical model, then the question arises of systematizing a large set of calculated criteria. Usually, one type of observables (for example, particle spectra) is analyzed separately from others. This leads to a contradicting interpretation of the experimental data. We came to conclusion that need the quantitative characteristics of the degree of agreement between theoretical models and experiments, which having a large amount of observational material, expressed by a single number for each model and for the each energy of heavy-ion collision. An attempt in this direction was made in [11], but with the averaging of the quality criteria $\chi^2$ of not compatible models, which somewhat smeared the final result.

This article is organized as follows. The second section provides a mathematical justification for the proposed method. In the third section, we apply it to published theoretical and experimental data with a graphical display of the result that unambiguously highlights the best theoretical model, and in fourth section we make some assertions. The fifth section contains the conclusion.

## 1. Theoretical justification of the method

Let ***A*** is a physical observable. A good criterion for the coincidence of some model $\boldsymbol{T_1}$ and experiment is the chi-square $\chi^2$:

$$\chi^2(A)_{T_1} = \sum_{i=1}^{n} \frac{(A_{\exp,i} - A_{th,i})^2}{\sigma_i^2}, \tag{1}$$

where $\sigma_i^2$ is the square of the experimental error of the physical observable $\boldsymbol{A_{exp,i}}$ measured in the i-th kinematic area, ***n*** is the number of data points of physical observable ***A***. Here, each physical observable, measured experimentally $\boldsymbol{A_{exp,i}}$ or calculated theoretically $\boldsymbol{A_{th,i}}$, is the result of averaging a series of measurements from thousands or millions of events - collisions of nuclei. The entire kinematic area in which the observable ***A*** is measured is divided into ***n*** non-overlapping areas. Therefore, the summation in (1) runs over the entire kinematic area of measurements of ***A***. Therefore, ***A*** is usually presented as a graph of ***n*** points or through a table.

Another criterion, rarely used in practice (but often in the laboratory practice of university courses), is the relative difference between the experimental value and theoretical prediction:

$$\delta(A)_{T_1} = \sum_{i=1}^{n} \left| \frac{A_{\exp,i} - A_{th,i}}{A_{\exp,i}} \right|. \tag{2}$$

Let the set of physical observables is $\boldsymbol{s_1} = \{A_1, ..., A_l\}$. After applying (1 - 2), sets of criteria can be obtained: $\{\chi^2(A_1)_{T_1}, ..., \chi^2(A_l)_{T_1}\}$ and $\{\delta(A_1)_{T_1}, ..., \delta(A_l)_{T_1}\}$. The next two values can be a quantitative expression of the degree of agreement between the model $\boldsymbol{T_1}$ and the experimental data set $\boldsymbol{s_1}$:

$$\chi^2(s_1)_{T_1} = \frac{\sum_{m=1}^{l}\chi^2(A_m)_{T_1}}{\sum_{m=1}^{l}n(A_m)}; \ \delta(s_1)_{T_1} = \frac{\sum_{m=1}^{l}\delta(A_m)_{T_1}}{\sum_{m=1}^{l}n(A_m)}, \tag{3}$$

where $\boldsymbol{n(A_m)}$ is the number of data points for the physical observable $\boldsymbol{A_m}$.

Now, in order to compare the $\boldsymbol{T_1}$ model with another set of experimental data $\boldsymbol{s_2}$ of physical observables $\{B_1, ..., B_k\}$ (related to other types of particles or physical processes), analogue (3) should be calculated:

$$\chi^2(s_2)_{T_1} = \frac{\sum_{m=1}^{k}\chi^2(B_m)_{T_1}}{\sum_{m=1}^{k}n(B_m)}; \ \delta(s_2)_{T_1} = \frac{\sum_{m=1}^{k}\delta(B_m)_{T_1}}{\sum_{m=1}^{k}n(B_m)}, \tag{4}$$

where $\boldsymbol{n(B_m)}$ is now the number of data points for the physical observable $\boldsymbol{B_m}$.

Comparing (3) with (4) in order to understand which set of observables the model $\boldsymbol{T_1}$ describes best of all, a problem arises. If all $\chi^2(A_m)_{T_1}$ or $\chi^2(B_m)_{T_1}$ ($\delta(A_m)_{T_1}$ or $\delta(B_m)_{T_1}$) have approximately the same value, then the sum (3) or (4) will lose some terms in the numerator that have the fewest data points. As a result, we lose some information about the physical processes under study, and we compare the truncated data sets. Moreover, using any weighted averaging, we reduce the set of observables, which distorts the analysis. To avoid this truncation of data, a modification has been made (1 - 4):

$$\left\langle \chi^2(A)_{T_1} \right\rangle = \frac{1}{n(A)}\sum_{i=1}^{n(A)}\frac{(A_{\exp,i} - A_{th,i})^2}{\sigma_i^2}, \tag{5}$$

$$\left\langle \delta(A)_{T_1} \right\rangle = \frac{1}{n(A)}\sum_{i=1}^{n(A)}\left|\frac{A_{\exp,i} - A_{th,i}}{A_{\exp,i}}\right|, \tag{6}$$

$$\left\langle \chi^2(s_1)_{T_1} \right\rangle = \frac{1}{l}\sum_{m=1}^{l}\left\langle \chi^2(A_m)_{T_1} \right\rangle; \ \left\langle \delta(s_1)_{T_1} \right\rangle = \frac{1}{l}\sum_{m=1}^{l}\left\langle \delta(A_m)_{T_1} \right\rangle, \tag{7}$$

$$\left\langle \chi^2(s_2)_{T_1} \right\rangle = \frac{1}{k}\sum_{m=1}^{k}\left\langle \chi^2(B_m)_{T_1} \right\rangle; \ \left\langle \delta(s_2)_{T_1} \right\rangle = \frac{1}{k}\sum_{m=1}^{k}\left\langle \delta(B_m)_{T_1} \right\rangle, \tag{8}$$

In formulas (7-8), we average criteria over the number of observables in each set of observables. Such averaging gives possibility for correct calculation of criteria for two sets inside one model $T_1$. To obtain criterion of comparison of the model $T_1$ with united sets $s_1$ and $s_2$, averaging of criteria over these sets is needed:

$$\left\langle \chi^2(s_1,s_2)_{T_1} \right\rangle = \frac{1}{2}\left(\left\langle \chi^2(s_1)_{T_1} \right\rangle + \left\langle \chi^2(s_2)_{T_1} \right\rangle\right), \tag{9}$$

$$\left\langle \delta(s_1,s_2)_{T_1} \right\rangle = \frac{1}{2}\left(\left\langle \delta(s_1)_{T_1} \right\rangle + \left\langle \delta(s_2)_{T_1} \right\rangle\right). \tag{10}$$

And for an arbitrary number $N$ of sets of observables, if $s = \bigcup_{i=1}^{N} s_i$ then:

$$\left\langle \chi^2(s)_{T_1} \right\rangle = \frac{1}{N}\sum_{i=1}^{N}\left\langle \chi^2(s_i)_{T_1} \right\rangle,$$
$$\left\langle \delta(s)_{T_1} \right\rangle = \frac{1}{N}\sum_{i=1}^{N}\left\langle \delta(s_i)_{T_1} \right\rangle. \tag{11}$$

Since each set of physical observables belongs to its own kinematic area, the arithmetic averaging of the criteria gives the final criterion which is uniformly distributed over the union of the kinematic areas of all sets of observables. Repeating the same analysis (5 – 11) for another model $T_2$ with respect to the same $N$ sets of physical observables makes it possible to compare criteria, for example, $\left\langle \chi^2(s)_{T_1} \right\rangle$ and $\left\langle \chi^2(s)_{T_2} \right\rangle$ of different models $T_1$ and $T_2$, that is, to compare the quality of theories in describing any amount experimental data.

## 2. Application to real experiments and theories

We take published results of three-fluid dynamic (3FD) model which uses three versions of equation of state (EoS) of nuclear matter created in heavy-ion collisions [12], [13]: $T_1$ is a 3FD model with 2-phase EoS, that is with first-order phase transition to deconfined state of nuclear matter, $T_2$ is a 3FD with EoS considering smooth crossover transition to deconfined state, $T_3$ is a 3FD with purely hadronic EoS. The 3FD model was applied to the experimental data for central heavy-ion collisions from AGS to RHIC energies $\sqrt{s_{NN}}$ = 2.7 GeV ÷ 62.4 GeV [14], [15].

We have applied formulas (6 – 8, 10, 11), concerning the relative criteria, to the following sets of physical observables: $s_1$ = ($Y^{\pi^\pm}$, $dY^{\pi^\pm}$); $s_2$ = ($Y^{p^+}$, $dY^{p^+}$); $s_3$ = ( $Y^{p^-}, dY^{p^-}$); $s_4$ = ($Y^{K^+}$, $dY^{K^+}$); $s_5$ = ($Y^{K^-}, dY^{K^-}$); $s_6$ = ($Y^{\Lambda^0}, dY^{\Lambda^0}$); $s_7$ = ($Y^{\Xi^+}$, $dY^{\Xi^+}$); $s_8$ = ($Y^{\Xi^-}, dY^{\Xi^-}$); $s_9$ = ($dY^{\Omega^-}$); $s_{10}$ = ($Y^{\bar{\Lambda}}$, $dY^{\bar{\Lambda}}$), $s_{11}$ = ($Y^{\phi}, dY^{\phi}$), where $\boldsymbol{Y^{particle}}$ is a total yield of given particle, calculated by taking integral from particle rapidity distributions ***dN/dy*** of [14], [15]. $\boldsymbol{dY^{particle}}$ is a midrapidity multiplicity $\left.\frac{dN}{dy}\right|_{y=0}$ of given particle taken from Fig. 9 of [15]. For all these physical observables, number of data points is one: $\boldsymbol{n_i = 1}$. Relative criteria are expressed as a percentage by multiplying the calculated values by 100 and the results are shown in Fig. 1.

At first, for charged particles, a separate averaging of relative criteria over each isospin group (***Q***) was done for protons: $\left\langle \delta(s_{p^\pm}) \right\rangle = \frac{\langle \delta(s_2) \rangle + \langle \delta(s_3) \rangle}{2}$, for kaons: $\left\langle \delta(s_{K^\pm}) \right\rangle = \frac{\langle \delta(s_4) \rangle + \langle \delta(s_5) \rangle}{2}$, for Ξ's: $\left\langle \delta(s_{\Xi^\pm}) \right\rangle = \frac{\langle \delta(s_7) \rangle + \langle \delta(s_8) \rangle}{2}$. Then sets of criteria for light flavor (***LF***) particles were averaged between themselves: $\left\langle \delta(s_{LF}) \right\rangle = \frac{\langle \delta(s_1) \rangle + \left\langle \delta(s_{p^\pm}) \right\rangle}{2}$. Analogously averaging inside strangeness group (***S***) was done also: $\left\langle \delta(s_S) \right\rangle = \frac{\left\langle \delta(s_{K^\pm}) \right\rangle + \left\langle \delta(s_{\Xi^\pm}) \right\rangle + \langle \delta(s_9) \rangle + \langle \delta(s_{11}) \rangle + \left( \langle \delta(s_6) \rangle + \delta(s_{10}) \right)/2}{5}$. Then two relative criteria for light flavor and strangeness are averaged between themselves: $\left\langle \delta(s_{Q,S,LF}) \right\rangle = \frac{\langle \delta(s_{LF}) \rangle + \langle \delta(s_S) \rangle}{2}$. Three numbers, $\left\langle \delta(s_{Q,S,LF})_{T_i} \right\rangle$ (where *i=1, 2, 3*), expressing quality of each version of 3FD model, were obtained for each energy of collision.

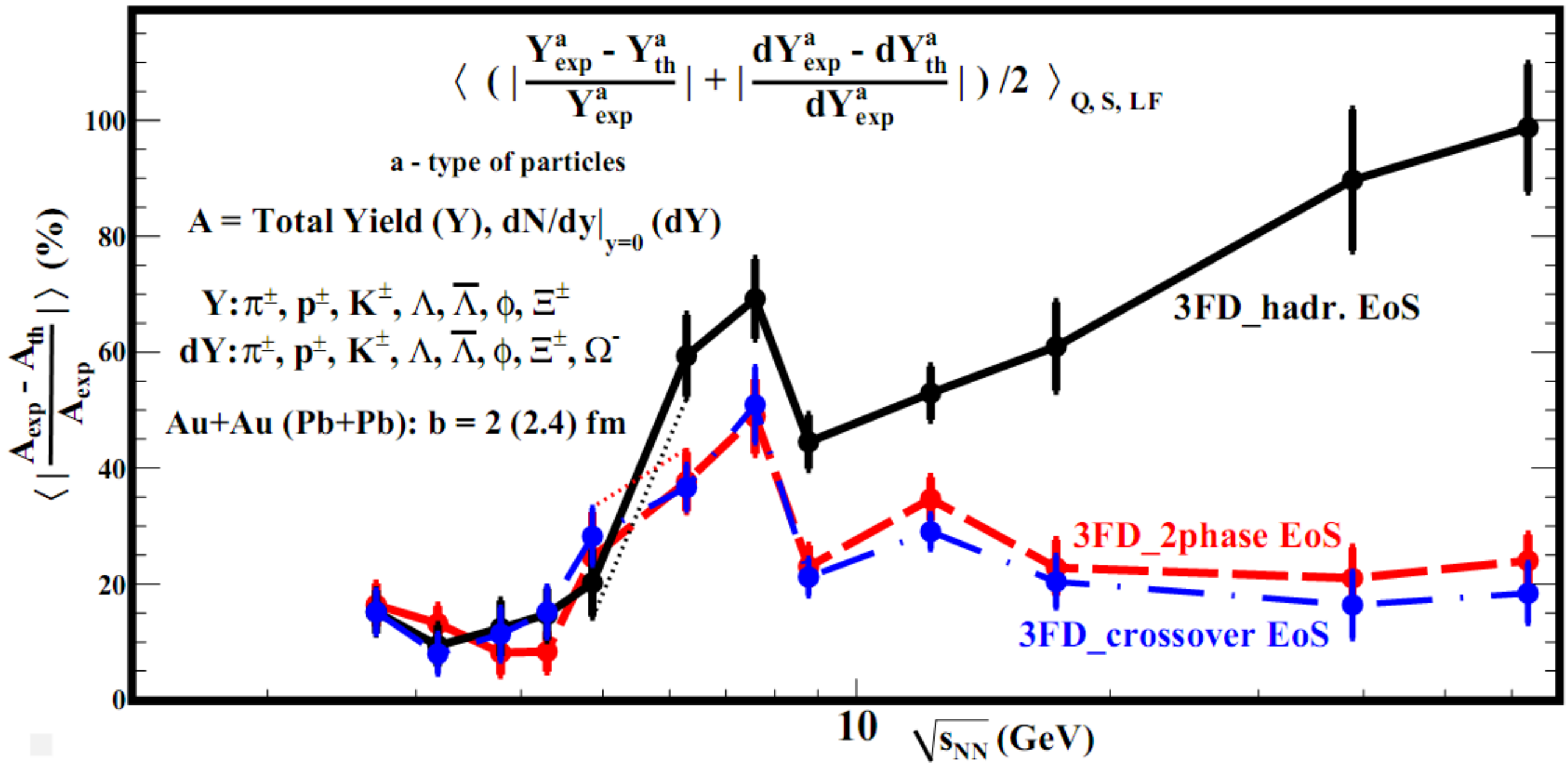

**Figure 1.** *Relative criteria of comparison between three versions of 3FD model and experimental data of total and mid-rapidity particle yields as function of the energy of central heavy-ion collisions. The formula at the top is demonstrating the method, described in the text. Symbols* ***Q, LF, S*** *are denoting the procedure of averaging inside groups of isospin, light flavor and strangeness, respectively, and final averaging between them (see text).*

The same procedure was done for directed flows $v_1(y)$ for protons, antiprotons and pions from mid-central heavy-ion collisions at energies $\sqrt{s_{NN}}$ = 2.7 ÷ 27 GeV, which were taken from Fig. 1-3 of [16]. Criteria were calculated by (5 – 11). Both types of criteria show similar behavior (Fig. 2). Relative criteria were no longer multiplied by 100. For each collision energy, the following sets of physical observables were taken: $s_1 = \left(\upsilon_1^{\pi^+}(y), \upsilon_1^{\pi^-}(y)\right), s_2 = \left(\upsilon_1^{p^+}(y), \upsilon_1^{p^-}(y)\right)$.

## 3. Discussion

It can be seen from Fig. 1 that all three versions of 3FD model are in poor agreement with experimental data in the central heavy ion collisions energy range of $\sqrt{s_{NN}} = 5 \div 9$ GeV. This may indicate that in this energy range the equation of state of nuclear matter has other parameters than those accepted in the 3FD model. If a phase transition of nuclear matter from the hadronic phase to the quark-gluon phase occurs somewhere, then at energies below $\sqrt{s_{NN}} = 5$ GeV and higher $\sqrt{s_{NN}} = 9$ GeV. Between these energies, nuclear matter is neither in a purely hadronic nor in a quark-gluon state.

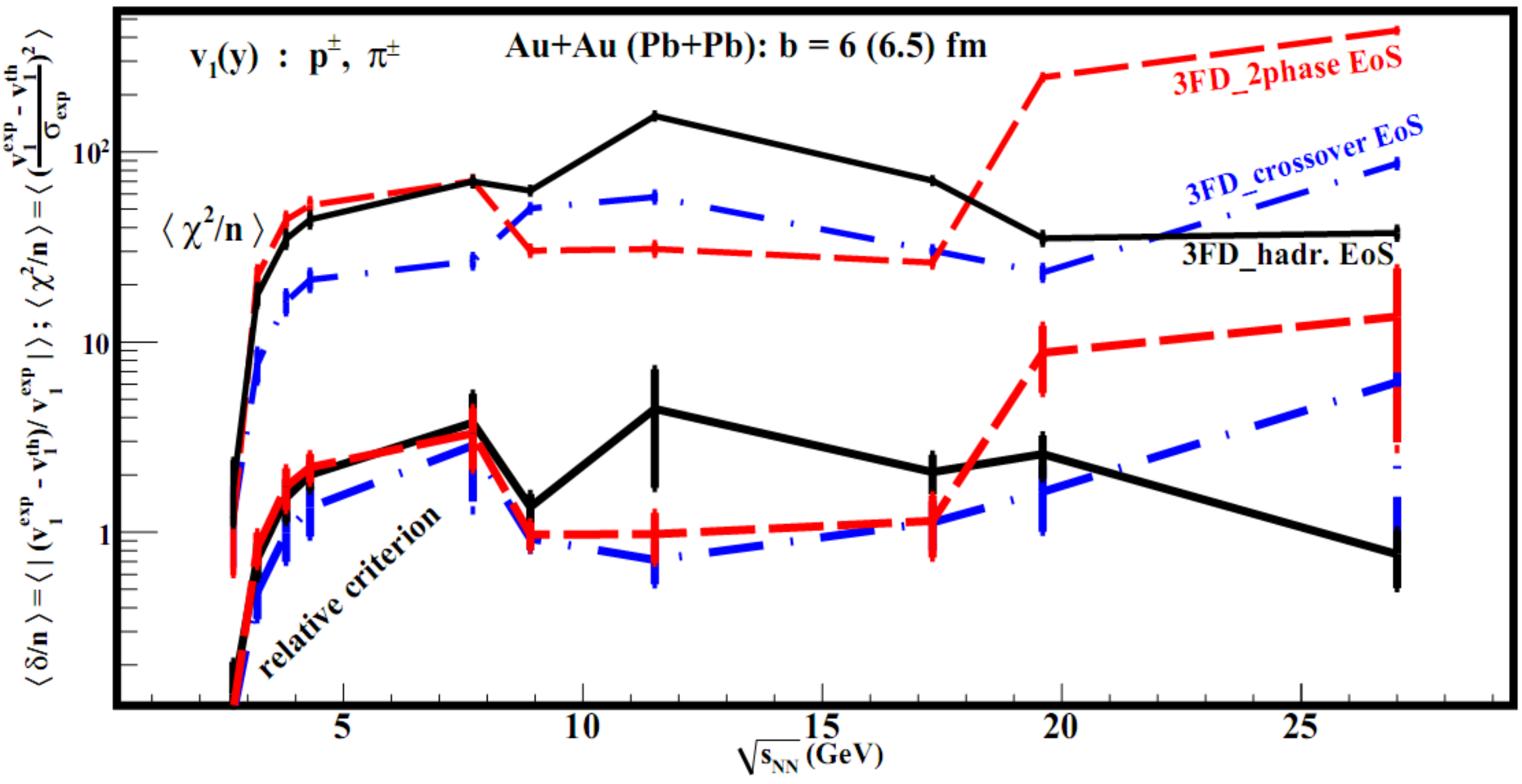

**Figure 2.** *Criteria for comparison between three versions of the 3FD model and experimental data of directed flows of light flavor particles depending on the energy of mid-central heavy ion collisions.*

In [17] it was shown that difference in the measurements of yields of hyperons by two experiments NA49 and NA57 is caused by quantum nature of the fireball created in heavy ion collisions, that is with probability around 50% occurs creation of fireball with ignited QGP state $|QGP\rangle$ at $\sqrt{s_{NN}}$ = 17.3 GeV in Pb+Pb central collisions, and with probability is around 50% we have fireball creation without igniting of QGP state, and we write this state in the form $|Hadronic\rangle$. Thus, we have suspicion that NA57 measures hyperon yields from mixture of two quantum states of fireball with predominance of $|QGP\rangle$ state, and NA49 measures hyperon yields from interference of two quantum states of fireball. This results in different measurement values in each experiment. Taking into account that each of the three versions of the 3FD model does not contradict physical laws, it can be assumed that there are three real scenarios for the evolution of the fireball in nature, that is, we have three quantum states of the fireball: $|Hadronic\rangle$, $|2phase\rangle$, $|crossover\rangle$, where the last two represent the $|QGP\rangle$ state via superposition. As a result, we must represent an arbitrary quantum state of fireball through a superposition of these three states, which is shown in Fig. 3. The amplitudes of these quantum states depend on the energy and centrality of heavy ion collisions. Thus, looking on Fig.1 we can assume that at energies $\sqrt{s_{NN}}$ > 17.3 GeV the amplitudes of the $|2phase\rangle$ and $|crossover\rangle$ states are almost constant and close to each other. But the amplitude of the $|Hadronic\rangle$ state decreases with increasing of energy of heavy ion collisions remaining non-zero even at $\sqrt{s_{NN}}$ = 63 GeV!

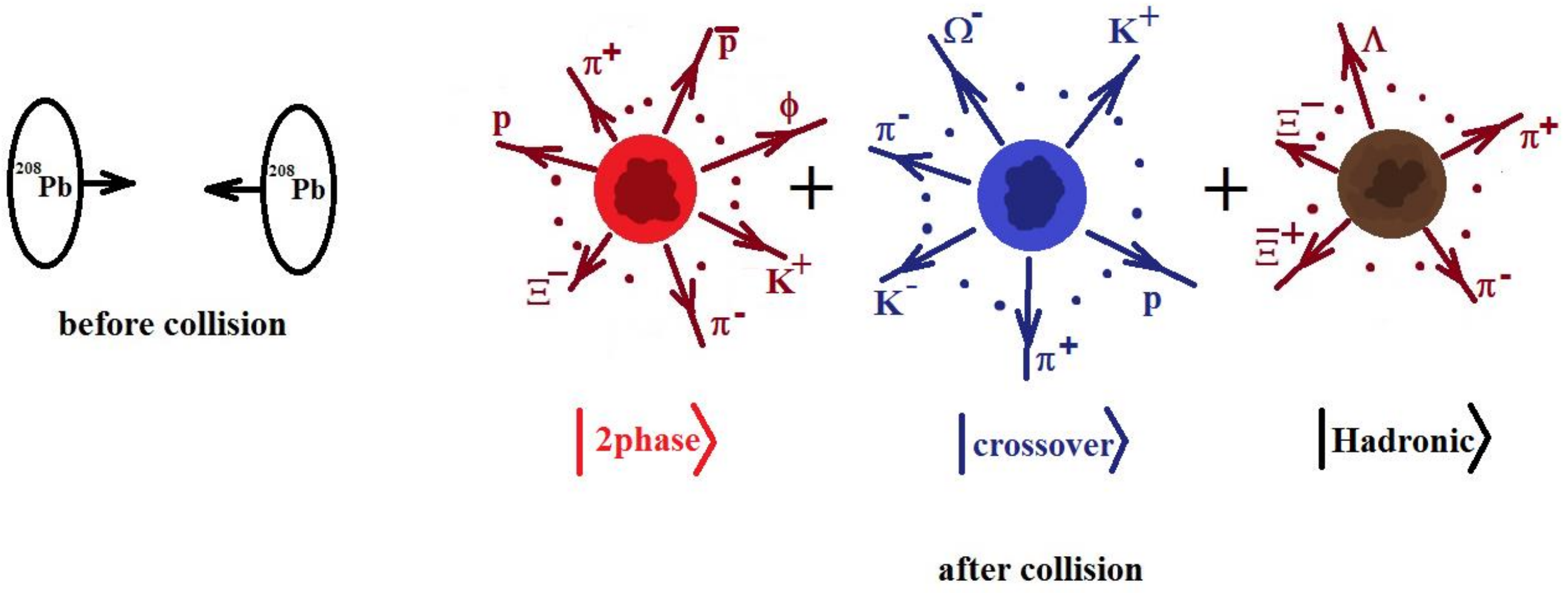

**Figure 3.** *Diagram of the superposition of three quantum states of a fireball formed in collisions of heavy ions at relativistic energies. The dots represent lines of other particles evaporating from the fireball.*

Probably, at collision energies of heavy ions of $\sqrt{s_{NN}}$ = 5 ÷ 9 GeV, there are other quantum states corresponding to the properties of nuclear matter not taken into account by the 3FD model, so the quantum states in the diagram in Fig. 3 will be different. This diagram in Fig. 3 relates to energies 2.7 < $\sqrt{s_{NN}}$ < 5 GeV and $\sqrt{s_{NN}}$ > 9 GeV. It is fundamental that the amplitudes of competing quantum states and representing a first-order phase transition $|2\,phase\rangle$ and a smooth transition to the partonic state $|crossover\rangle$ of nuclear matter, both of these amplitudes do not vanish in the same energy range!

Large values of the criteria for directed flows in Fig. 2 can be attributed to bad statistics, since we have only two types of particles - (anti)protons and charged pions. The suspiciously better agreement hadronic version of 3FD model at energies $\sqrt{s_{NN}}$ > 20 GeV is explained in [18] by an incorrect choice of the parameters of deconfined nuclear matter.

## 5. Conclusion

The shown method of comparing theoretical predictions with a large set of experimental data provides a clear opportunity to assess the quality of the theory and choose the best theory among many theoretical models. In experiments on heavy ions, a large number of physical observables are measured for different types of elementary particles. Each of the existing theoretical models describes the experiment well only in a narrow range of experimental data, competing with other models almost on an equal footing in the rest of the measurement region. In the vast majority of cases, even a comparison of several dozen experimental spectra with theoretical calculations is carried out by eye, which makes it impossible to make an unambiguous conclusion about the quality of theories and the choice of the best of them. The method proposed in this article is extremely simple - after grouping the data and calculating the appropriate criterion for each spectrum, it is necessary to average the obtained criteria. Further visualization of the data in the form of, for example, the distribution of the average criterion over energy, gives an unambiguous idea of the quality of the model over the entire measurement range.

At the same time, when exploring a quantum object, we must understand that nature is richer in the manifestation of physical phenomena than the human imagination. Competing models that assume different evolution of a physical object might be represented in nature as different quantum states of an object that are realized under the same conditions. Thus, taking into account the quantum nature of the fireball, we can assume that even at energy of $\sqrt{s_{NN}}$ = 63 GeV of central heavy ion collisions, there are such events where the QGP does not ignite and during lifetime the fireball consists only of pure hadronic matter (not deconfined). The probability of such an event is small, but not zero. In the experiment, this is not taken into account and it is assumed by an unspoken, unsupported agreement that such events do not happen at such energy. On the other hand, at an energy of central heavy-ion collisions of around $\sqrt{s_{NN}}$ = 12 GeV, the number of events without QGP ignition can be comparable to the number of events with QGP ignition. The ignition of the QGP itself can proceed according to two scenarios - with a first-order phase transition and through a smooth crossover. And again, according to an unspoken agreement, not supported by anything, it is believed that only one scenario of the process is realized in nature. The quantum nature of the fireball is somehow completely ignored. At the same time, the question of the possibility of applying an appropriate trigger in an experiment for selecting events with different fireball quantum states remains open.